\documentclass[twocolumn,
aps,prl,twoside,superscriptaddress,floatfix
]{revtex4-1}

\usepackage{dcolumn}
\usepackage{float}
\usepackage{bm}
\usepackage[T1]{fontenc}
\usepackage{amsmath}
\usepackage{graphicx}
\usepackage{tabularx}
\usepackage{multirow}
\usepackage{rotating}
\usepackage{makecell}
\usepackage{hyperref}
\hypersetup{colorlinks=true, linkcolor=blue, citecolor=blue, urlcolor=blue}
\usepackage{tikz}
\usepackage[normalem]{ulem}
\usepackage{pgfplots}

\begin{document}

\title{Global formation of topological defects in the multiferroic hexagonal manganites}

\author{Q. N. Meier$^{\ast}$}
\affiliation{Department of Materials, ETH Zurich, 8093 Z{\"u}rich, Switzerland}
\author{M. Lilienblum$^{\ast}$}
\affiliation{Department of Materials, ETH Zurich, 8093 Z{\"u}rich, Switzerland}
\author{S. M. Griffin}
\affiliation{Department of Physics, University of California Berkeley, Berkeley, CA 94720, USA}
\affiliation{Molecular Foundry, Lawrence Berkeley National Laboratory, Berkeley, CA 94720, USA}
\author{K. Conder}
\affiliation{Laboratory for Scientific Developments and Novel Materials, Paul Scherrer Institute, 5232 Villigen, Switzerland}
\author{E. Pomjakushina}
\affiliation{Laboratory for Scientific Developments and Novel Materials, Paul Scherrer Institute, 5232 Villigen, Switzerland}
\author{Z. Yan}
\affiliation{Materials Sciences Division, Lawrence Berkeley National Laboratory, Berkeley, CA 94720, USA}
\affiliation{Department of Physics, ETH Zurich, Otto-Stern-Weg 1, 8093 Zürich, Switzerland}
\author{E. Bourret}
\affiliation{Materials Sciences Division, Lawrence Berkeley National Laboratory, Berkeley, CA 94720, USA}
\author{D. Meier}
\affiliation{Department of Materials Science and Engineering, Norwegian University of Science and Technology, 7491 Trondheim, Norway}
\author{F. Lichtenberg}
\affiliation{Department of Materials, ETH Zurich, 8093 Z{\"u}rich, Switzerland}
\author{E. K. H. Salje}
\affiliation{Department of Earth Sciences, University of Cambridge, Cambridge CB2 3EQ, UK}
\author{N. A. Spaldin}
\affiliation{Department of Materials, ETH Zurich, 8093 Z{\"u}rich, Switzerland}
\author{M. Fiebig}
\affiliation{Department of Materials, ETH Zurich, 8093 Z{\"u}rich, Switzerland}
\author{A. Cano}
\affiliation{Department of Materials, ETH Zurich, 8093 Z{\"u}rich, Switzerland}
\affiliation{CNRS, University of Bordeaux, ICMCB, UPR 9048, 33600 Pessac, France}
\date{\today}

\begin{abstract}

The spontaneous transformations associated with symmetry-breaking phase transitions generate domain structures and defects that may be topological in nature. The formation of these defects can be described according to the Kibble-Zurek mechanism, which provides a generic relation that applies from cosmological to interatomic lengthscales. Its verification is challenging, however, in particular at the cosmological scale where experiments are impractical. While it has been demonstrated for selected condensed-matter systems, major questions remain regarding e.g. its degree of universality. Here we develop a global Kibble-Zurek picture from the condensed-matter level. We show theoretically that a transition between two fluctuation regimes (Ginzburg and mean-field) can lead to an intermediate region with reversed scaling, and we verify experimentally this behavior for the structural transition in the series of multiferroic hexagonal manganites. Trends across the series allow us to identify additional intrinsic features of the defect formation beyond the original Kibble-Zurek paradigm.

\end{abstract}

\maketitle

Topological defects are ubiquitous in nature, emerging in various forms in a large variety of physical systems from atomic to cosmic length scales. In the context of cosmology, Kibble first inspected the link between the possible topology of the corresponding defects and gauge symmetry breaking \cite{Kibble1976}. Subsequently, Zurek derived a scaling law relating the density of defects and the speed at which the transition point is crossed \cite{Zurek1985}. Their combined theory is known as the Kibble-Zurek mechanism.  Under the appropriate conditions, this mechanism is expected to describe the formation of topological defects in a system that is driven through a continuous phase transition at a finite cooling rate. Since Kibble-Zurek scaling is determined by the critical behavior and should be the same for all systems in the same universality class, Zurek proposed the study of condensed-matter analogues to cosmic systems for its verification.

A variety of condensed matter systems have been investigated to date in an effort to verify the Kibble-Zurek mechanism. Early attempts were carried out on liquid crystals \cite{Chuang91,Ducci1999}, superfluid $^4$He and $^3$He \cite{Hendry1994,Dodd1998,Baeuerle1996,Ruutu1996} and superconducting rings \cite{Carmi00,Monaco2009}. More recent studies have been conducted on multiferroics \cite{Griffin2012,Chae2012,Lin2014}, Bose-Einstein condensates \cite{Su2013,{donadello16}}, ionic crystals \cite{Ulm13,pyka13}, Landau-Zener setups \cite{Xu14}, and colloidal monolayers \cite{Deutschlander15}; for a review see Ref.~\cite{DelCampo2014}. The case of multiferroics is particularly interesting, as they have provided the first experimental setting clearly compatible with a Kibble-Zurek scaling beyond mean-field \cite{Griffin2012}. On the other hand, for the same system a drastic reversal of this scaling  (termed ``anti-Kibble-Zurek scaling'' \cite{Griffin2012}) has been reported for fast quenches, although its origin is not understood and its existence has been questioned \cite{Lin2014}.

In this work we combine first-principles calculations and the theory of critical phenomena to provide a global picture of the Kibble-Zurek mechanism in which, by increasing the cooling rate, defect formation evolves from the fluctuation-dominated Ginzburg region to the mean-field regime. This picture naturally encompasses features of anti-Kibble-Zurek behavior, which can emerge from the crossover between these two distinct regimes. Our model system for this investigation is the series of hexagonal multiferroic manganites, $R$MnO$_3$, here with $R = $ Y, Dy, Er and Tm. 

In our scanning probe measurements, both the Kibble-Zurek scaling and the anti-Kibble-Zurek behavior are demonstrated unequivocally as a general feature in hexagonal manganites. In addition, trends that we uncover by studying the $R$MnO$_3$ series as a whole reveal additional quantitative features suggesting that the topological defect formation is affected by supplementary ingredients beyond the original Kibble-Zurek theory. 

We discuss emergence of additional time- and length-scales as likely candidates for these extra features, which can appear naturally from the propagation of the phase-transition front, the vortex-growth process, or directly from the eventual discrete nature of the corresponding symmetry breaking.

\begin{figure}[t!]
\includegraphics[width=.4\textwidth,keepaspectratio=true]{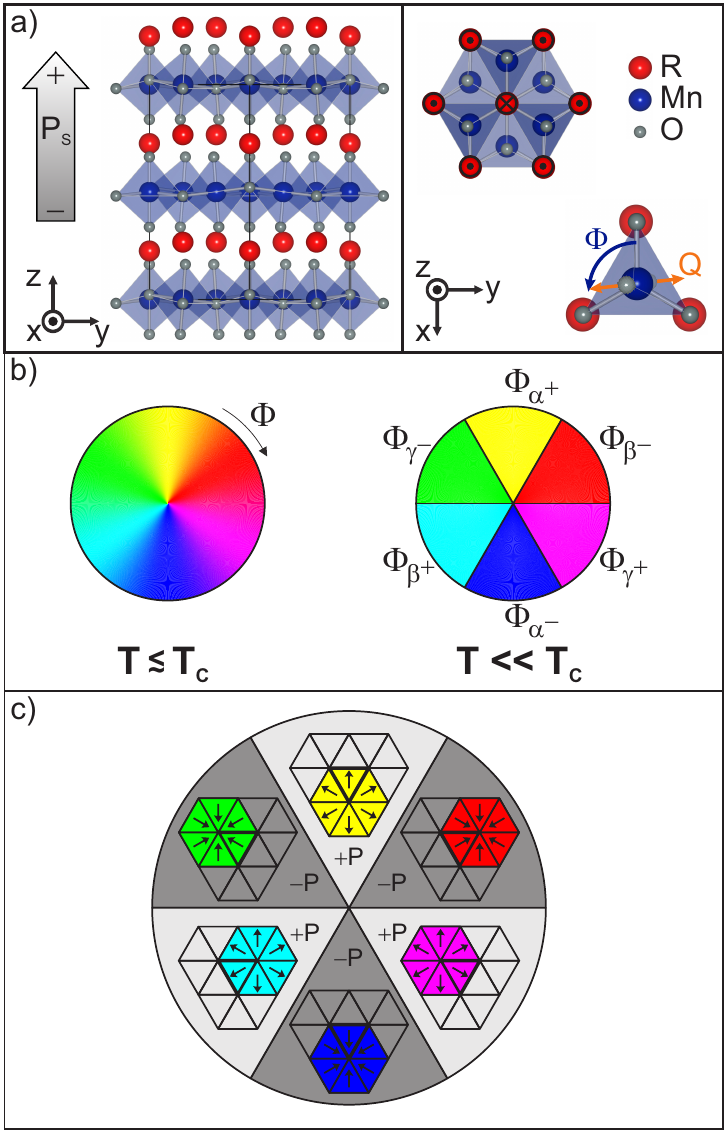}
\caption{Formation of topological defects in the hexagonal manganites. {\bf a}, Side and top views of the unit cell showing the arrangement of the tilted MnO$_5$ bipyramids. At the ordering temperature $T_c$, sets of three bipyramids tilt towards or (in this case) away from a common center. This trimerizes the unit cell and induces a spontaneous polarization $\pm P$. The order parameter of the trimerization-polarization is $(Q\cos\Phi, Q\sin\Phi)$ with $Q$ and $\Phi$ as sketched. {\bf b}, Topological defects are lines (points on the sample surface) around which $\Phi$ changes monotonically in a clockwise or counterclockwise fashion. Close to $T_c$ this change is gradual due to the ``dangerously irrelevant'' character of the $Z_6$ anisotropy. At lower temperature the $Z_6$ anisotropy becomes fully relevant and six domain states with discrete values $\Phi=n\cdot60^{\circ}$, $n=0,1,\dots, 5$, emerge. {\bf c}, Possible arrangement of the six domain states around the vortex-like topological defect. For each domain state the bipyramidal tilt pattern is indicated with arrows representing $Q$ and $\Phi$.}
\label{fig:lhs}
\end{figure}

We prepared single crystals of YMnO$_3$, DyMnO$_3$, and ErMnO$_3$ using the floating-zone (FZ) technique as described in Methods (see also Supplementary Information). To complete our analysis, we also consider data for TmMnO$_3$ reported in \cite{Lin2014}. These $R$MnO$_3$ compounds undergo a high-temperature lattice-distortive unit-cell-trimerizing transition at $T_c^\text{Y}\simeq$ 1259~K , $T_c^\text{Dy}\simeq$ 1223~K, $T_c^\text{Er} \simeq $ 1429~K \cite{Chae2012}, and $T_c^\text{Tm} \simeq $1514~K \cite{Lilienblum2015,Chae2012}. By driving the systems through this structural transition, topological defects are created in the peculiar form of discrete vortices as sketched in Fig.~\ref{fig:lhs} \cite{Griffin2012,Chae2012,Lin2014,Choi2010,Zhang2013,Artyukhin2014,Lilienblum2015}. These vortices correspond to particular solutions of the Landau free energy \cite{Artyukhin2014}
\begin{equation}\label{lfe}
  F={a\over 2}Q^2 + {b\over 4}Q^4 + {1\over 6}(c + c'\cos 6 \Phi)Q^6 + {s\over 2}[(\nabla Q)^2 + Q^2(\nabla \Phi)^2],
\end{equation}
where $\mathbf Q = (Q \cos \Phi, Q \sin \Phi)$ is the primary order parameter associated with the condensation of a zone-boundary $K_3$ phonon. This condensation induces the spontaneous polarization $P \sim Q^3 \cos 3 \Phi$ ($\Phi = n \pi /3$, with $n = 0, 1, \dots, 5$). The polarization alternation of the resulting six trimerization-polarization domain states around the vortices (see Fig.~\ref{fig:lhs}c) enables their real space imaging by piezoresponse force microscopy (PFM). The parameters of Eq.~\ref{lfe} for the series of hexagonal manganites calculated in this work using density functional theory (see Methods) are given in Table~\ref{phyprop}.

\begin{table}[b!]
\setlength{\tabcolsep}{5pt} % Default value: 6pt
\centering
\begin{tabular}{l c c c c}
\hline \hline
& YMnO$_3$ & DyMnO$_3$ & ErMnO$_3$ & TmMnO$_3$  \\
\hline
$a_0$ (eV/A$^2$)& -3.2 & -3.6 & -3.8 & -3.9 \\
$b$ (eV/A$^4$) & 5.6 & 5.8& 6.1 &5.6 \\
$s_{xy}$ (eV) & 4.57 & 4.47 & 4.73 & 4.52 \\
$s_{z}$ (eV) & 17.2 & 18.7 & 18.9 & 20.18\\
\hline
$\xi_0$ (\AA) & 1.48 & 1.50  & 1.41   &  1.40  \\
$\tilde \xi_{0}$ (\AA) & 2.00 & 2.00 &  1.90  & 1.89  \\
$Gi$ & 0.27 & 0.24 & 0.28 &  0.27    \\  	
\hline \hline
\end{tabular}\caption{Parameters %$a_0$,$b$,$s_{xy}$,$s_z$ 
of the Landau free energy \eqref{lfe} obtained from DFT calculations (see Methods). Here $a_0$ denotes the zero-temperature value of the parameter $a$ (in the simplest case $a = -a_0 T_c\varepsilon$, where $\varepsilon = (T-T_c)/T_c$ is the reduced temperature). The parameter $s$ in \eqref{lfe} is the mean value of the anisotropic stiffness $s=(s_{xy}^2s_z)^{1/3}$. The zero-temperature correlation length $\xi_0$, renormalized zero-temperature correlation length $\tilde \xi_0$, and Ginzburg number $Gi$ are derived from these values.}\label{phyprop}
\end{table}

Our samples were annealed above the transition temperature and then quenched at cooling rates ranging from $10^{-2}$~K/min to $\approx 10^{5}$~K/min in a temperature interval around $T_c$ of at least $\pm 100$~K. To exclude surface-related effects, samples were then thinned by about 100~$\mu$m before polishing them and resolving the ferroelectric domain structure by PFM at room temperature (see Methods). The resulting images are shown for DyMnO$_3$ and ErMnO$_3$ in Fig.~\ref{VortexQuench}; the behavior of YMnO$_3$ is similar. Both compounds exhibit the characteristic domain pattern in which meeting points of the six trimerization-polarization domains identify the location of the topological vortex defects. DyMnO$_3$ shows an increase of vortex density $n$ up to a cooling rate of 1~K/min, followed by a striking decrease of $n$ by two orders of magnitude upon further increase of the cooling rate. ErMnO$_3$ displays qualitatively the same behaviour, but the decrease of $n$ sets in at higher cooling rates than in DyMnO$_3$. Note that previous experiments \cite{Chae2012,Lin2014} were performed at slower cooling rates, and so did not discover the turn around in the slope of $n$.

\begin{figure}[t!]
\includegraphics[width=.41\textwidth]{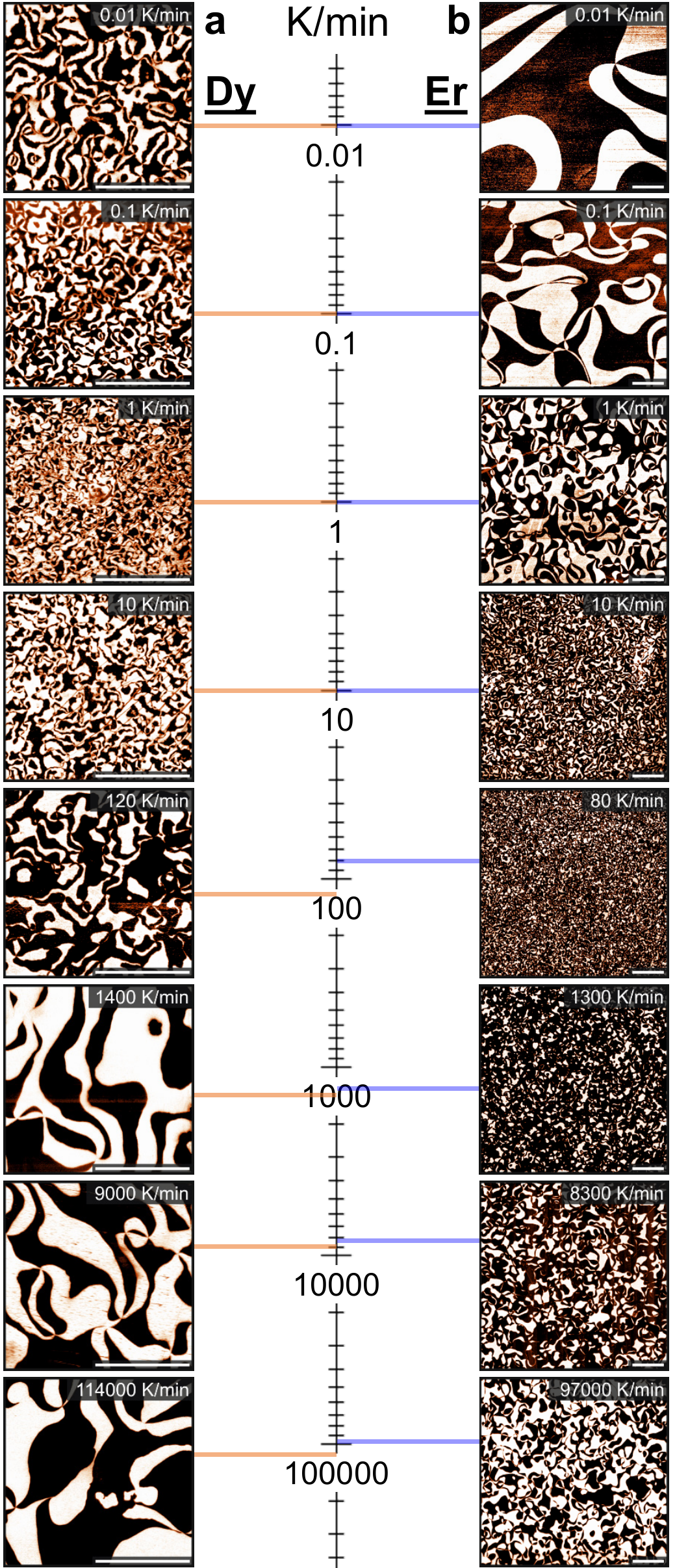}
\end{figure}
\begin{figure}[t!]
\caption{Spatial maps of the vortex-like ferroelectric domain pattern in hexagonal manganites for a range of different cooling rates through $T_c$. Images were recorded by PFM under ambient conditions. Labels denote the respective cooling rates. Scale bar in all images is 5~$\mu$m. {\bf a}, DyMnO$_3$. {\bf b}, ErMnO$_3$. The Kibble-Zurek-like increase of vortex density with cooling rate, followed by a reversal and decrease of vortex density is clear in both compounds.}
\label{VortexQuench}
\end{figure}

\begin{figure}[tb]
\includegraphics[width=.35\textwidth,keepaspectratio=true]{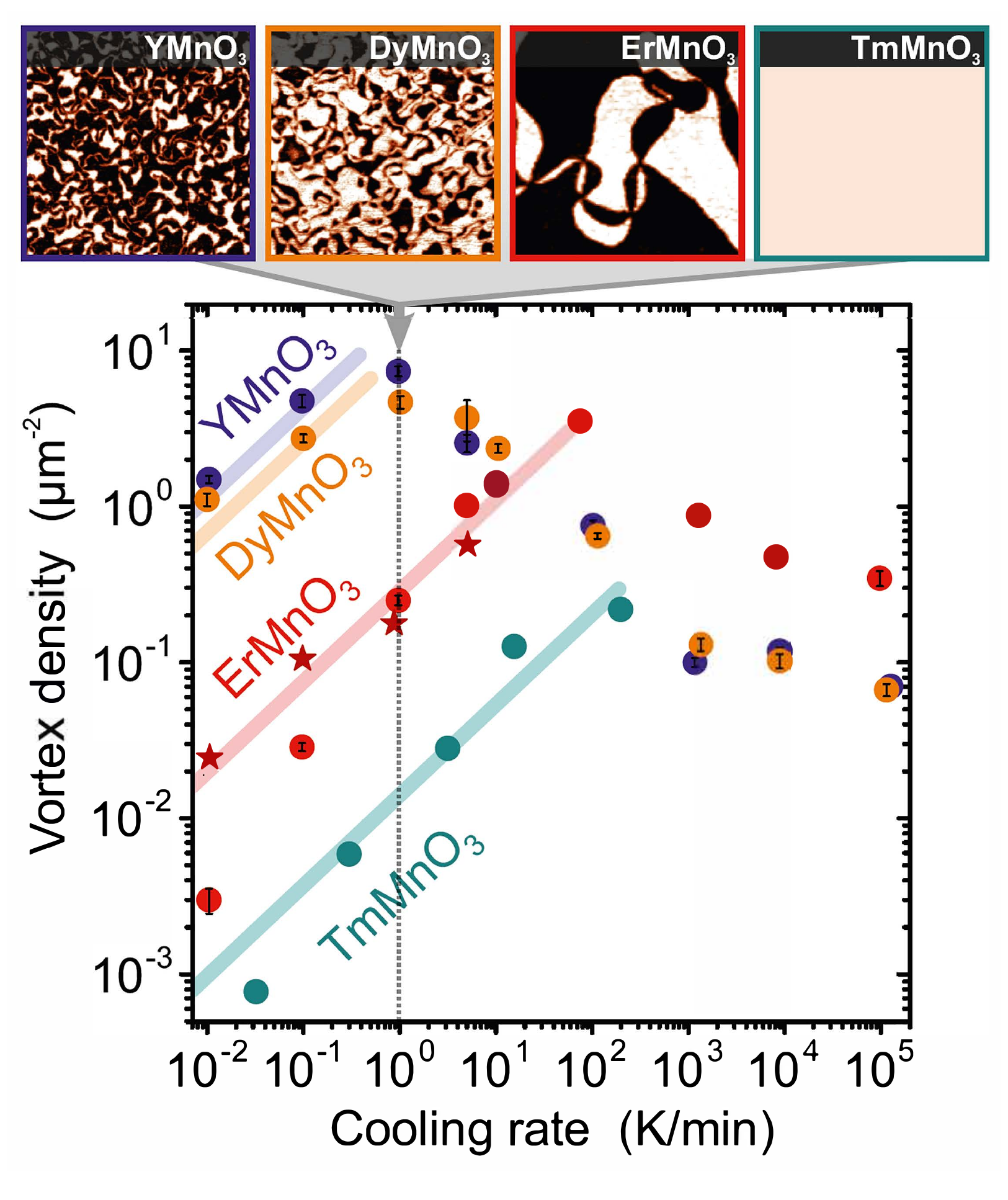}
\caption{Dependence of domain vortex density on cooling rate through $T_c$ across the $R$MnO$_3$ series, for $R$ = Y, Dy, Er, Tm. Two regimes, one in which the vortex density increases and one it which it decreases with cooling rate are obvious. Lines show the fitted Kibble-Zurek behaviour. Error bars represent the statistical error of the counted vortex number. Data for TmMnO$_3$ were taken from Ref.~\cite{Lin2014}. Starlike red symbols indicate data points taken on flux-grown samples in Ref.~\cite{Chae2012}. Insets show PFM images of the domain structure after quenching at 1~K/min for $5\times 5$~$\mu$m$^2$ sections. The TmMnO$_3$ image is sketched as its domain size would exceed the shown section.}
\label{VortexSummary}
\end{figure}

The vortex density as a function of cooling rate for all four $R$MnO$_3$ compounds is displayed in Fig.~\ref{VortexSummary}, clearly revealing that both the Kibble-Zurek and the anti-Kibble-Zurek behaviour are generic features of the vortex formation in the hexagonal magnanites. We also find an intriguing dependence on the crystal chemistry, with larger  $R^{3+}$ radius correlating with higher $n$ at a given cooling %rate and a decrease 
rate within the Kibble-Zurek region, as well as with a decrease of the cooling rate at which the turn around occurs. In addition, we observe deviations from Kibble-Zurek scaling in the ultraslow-cooling regime. We return to these deviations later, after discussing first the origin of the turn around between Kibble-Zurek and anti-Kibble-Zurek behavior.

According to the Kibble-Zurek mechanism, the vortices are expected to emerge from critical fluctuations with a density that is essentially determined by the rate of cooling through the phase transition and the critical slowing down of the system \cite{landau-khalatnikov}. When the relaxation of the order parameter becomes slower than the changes introduced by the quenching the thermal fluctuations freeze out and give rise to a non-equilibrated order-parameter distribution.

We first revise this picture by making a distinction between two fluctuation regimes, namely the Gaussian approximation about mean-field and the Ginzburg (or scaling) regimes \cite{landau2013v5,larkin2005,strukov1998}. Roughly speaking, fluctuations are assumed to be non-interacting fluctuations in the Gaussian approximation while their interaction becomes crucial and controls the critical properties in the Ginzburg regime. These fluctuations determine the vortex density $n$ expected according to the Kibble-Zurek hypothesis as 
\begin{align}\label{n}
  n \sim {1\over \xi ^{2}(t_*)} = {1 \over \xi^2_0}\Big({\tau_0\over \tau_q}\Big)^{2\nu/(1+z \nu)} .
\end{align}
%
%by determining the critical exponents, $\nu$ and $z$.
Here $\xi(t_*)$ is the coherence length at the freeze-out time $t_*$, which in turn is the time at which the quenching process becomes faster than the relaxation of the order parameter, $\tau$. For a linear quench with $T(t) =(1 + t /\tau_q) T_c$, where $\tau_q$ is the characteristic time set by the cooling rate $r=T_c\tau_q^{-1}$, the freeze-out time is given by
$t_* \sim (\tau_0 \tau_q^{z \nu})^{1/(1+z \nu)}$ \cite{Zurek1985,DelCampo2014}. 
Here $z$ and $\nu$ are the dynamical critical exponent and the critical exponent for the correlation length according to 
$\tau(\varepsilon) = \tau_0/|\varepsilon|^{z\nu}$ and $\xi(\varepsilon) = \xi_0 /|\varepsilon|^\nu$, where $\varepsilon = {(T - T_c)/ T_c}$, is the (time-dependent) reduced temperature \cite{Hohenberg15,salje92,salje88}. 

We see in Eq.\ \eqref{n} that the precise form of the Kibble-Zurek scaling is fundamentally related to the nature of the critical fluctuations at the freeze-out. In the mean-field regime the critical exponents are $\nu = 1/2$ and $z=2$, which leads to a mean-field Kibble-Zurek exponent ${2\nu/(1+z \nu)} = 1/2$. In addition, the microscopic correlation length is $\xi_0 = \sqrt{s/a_0}$ in terms of the Landau free-energy parameters in Eq.\ \eqref{lfe} with the expansion $a=-a_0T_c\varepsilon$. On the other hand, the critical exponents for Eq.\ \eqref{lfe} in the Ginzburg regime take the values $\nu = 0.672$ and $z \simeq 2$ of the 3D XY model \cite{nelson-prb-76,oshikawa-prb-00,Lin2014,Lilienblum2015}. 
As a result, the Kibble-Zurek exponent becomes ${2\nu/(1+z \nu)} =0.58$.  
The onset of interaction between fluctuations characterizing the transition to the Ginzburg regime modifies not only the critical exponents, but also ``microscopic'' parameters such as $\xi_0$, which becomes
\begin{align}\label{renormalized-xi0}
 {\tilde \xi}_0 = \Big({5 b k_B T \over \pi^2 \bar s^2 }\xi_0\Big)^{2\nu/5} \xi_0,
\end{align}
where $\bar s = (s_{xy}^2s_z)^{1/3}$ represents an averaged gradient coefficient (see Supplemental Material). 

The crossover between the mean-field and Ginzburg regimes is defined by the Ginzburg-Levanyuk criterion as the point at which the order-parameter fluctuations in a correlation volume of size  $\xi^3$ reach the magnitude of the order parameter itself \cite{landau2013v5,larkin2005,strukov1998}. This crossover can be estimated from observables such as the specific heat. Thus, with Eq.\ \eqref{lfe} one obtains the so-called the Ginzburg-Levanyuk number $Gi \equiv \varepsilon_\text{crossover} = (b k_B T_c)^2 / [(4 \pi)^2 a_0 \bar s^3]$ \cite{larkin2005}.

This has to be compared with the reduced freeze-out temperature $\varepsilon(t_*)=(\tau_0/\tau_q)^{1/(1+z\nu)}$ to determine the formation of vortices in the Kibble-Zurek picture. At slow cooling rates, where $\varepsilon(t_*)< Gi$, the Kibble-Zurek mechanism probes the fluctuation-dominated Ginzburg region. However, as the cooling rate increases, the freeze-out eventually occurs so far from $T_c$ that $\varepsilon(t_*)> Gi$ and hence the Kibble-Zurek physics emerges from Gaussian fluctuations in the mean-field regime. The critical exponents and the microscopic parameters then change accordingly.

\begin{figure}[tb]
\includegraphics[width=.45\textwidth]{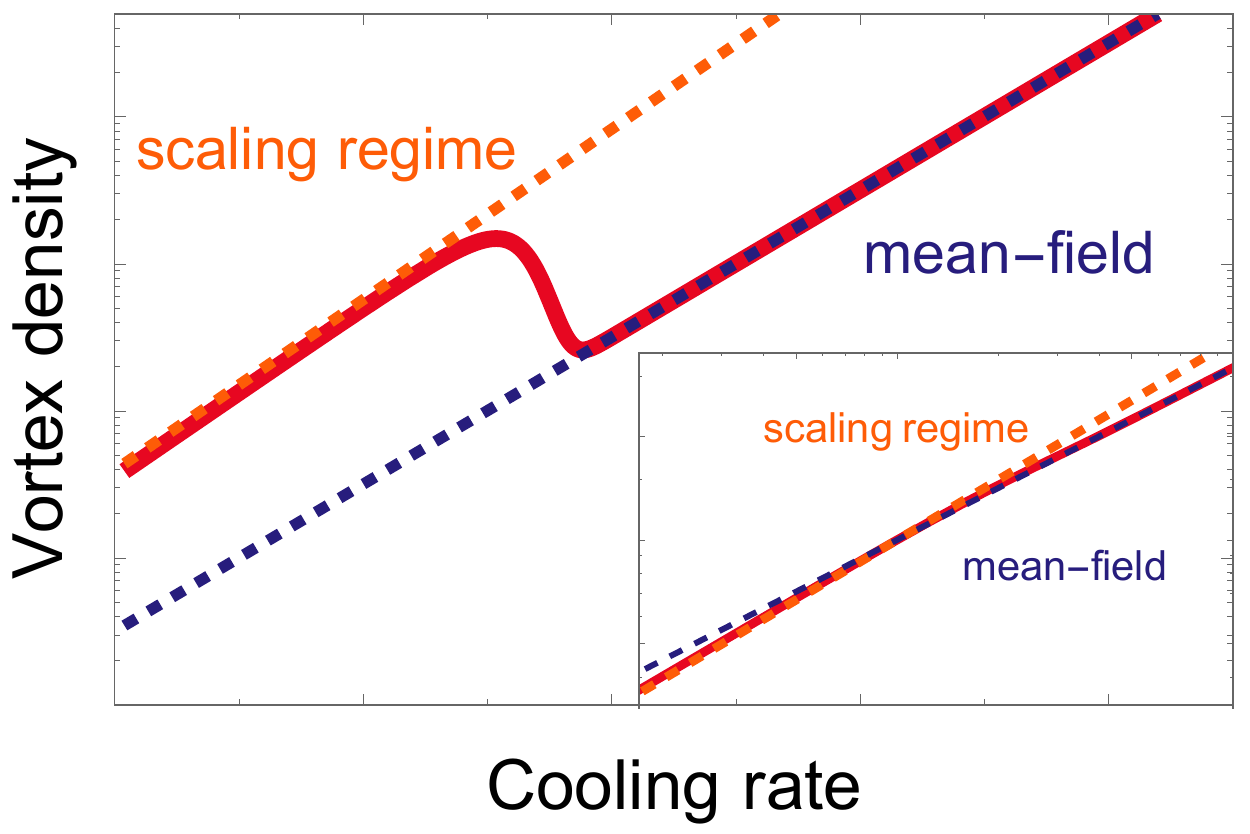}
\caption{Evolution of the Kibble-Zurek scaling. As the fluctuation regime probed in the experiment changes from Ginzburg (red dashed line) to mean-field (blue dashed line) behaviour, the density of defects can display either a dropdown behavior (red line with non-crossing of red and blue dashed lines) or a smooth transition (orange line with crossing of red and blue dashed lines, see inset). The scenario depends on the microscopic parameters of the system, foremost the renormalization of the relaxation time $\tau_0$ when entering the Ginzburg regime (see text).
}\label{fluct-to-mf}
\end{figure}

The expected crossover for a system described by Eq.\ \eqref{lfe} is illustrated in Fig.\ \ref{fluct-to-mf}. Since the Kibble-Zurek exponent $2\nu/(1+\nu z)$ is larger in the Ginzburg regime than in the mean-field regime an offset between the related lines in the double-logarithmic plot is expected. The density of defects is then expected to show a dropdown (red line in Fig.\ \ref{fluct-to-mf}) when the transition from the Ginzburg to the mean-field behaviour occurs. 
The $Gi$ values that we obtain from our density functional theory (DFT) calculations, given in Table~\ref{phyprop}, are compatible with our observed crossovers, further justifying our use of the Ginzburg regime value 0.58 for the Kibble-Zurek exponent in the scaling regime at slow cooling. 
We also see in Table \ref{phyprop}, however, that the hexagonal manganites have a consistently larger correlation length $\widetilde \xi_0$ in the Ginzburg than in the mean-field regime, $\xi_0$. This should tend to reduce the offset shown in Fig.~\ref{fluct-to-mf} and could even lead to a crossing of the Ginzburg and mean-field graphs (Fig.\ \ref{fluct-to-mf} inset), along with a smooth transfer between the regimes as indicated by the orange line. A central factor discriminating between the two scenarios, however, is the renormalization of the relaxation time $\tau_0$.  Equation (\ref{renormalized-xi0}) shows that the correlation length increases in the Ginzburg regime, and with this we expect the relaxation time $\tau_0$ to go up. Hence, the dropdown scenario is the more likely one, providing an explanation for our experimental results in Figs.\ \ref{VortexQuench} and \ref{VortexSummary}. We expect that experiments at higher quench rates, which we have not yet been able to access, should show a second turn around, as vortex densities begin once again to increase with quench rate following a mean-field scaling.

It is worth noting that, in the seemingly unrelated case of a quantum phase transition driven by a noisy control parameter, the possibility of a dropdown has also been pointed out \cite{dutta16}. This possibility is in fact analogous in the sense that the departure from the Kibble-Zurek scaling is also related to the running of the critical exponents --from their nominal value to the noise-fluctuation limit in that case.

Next we address the deviations observed in our ErMnO$_3$ samples in the ultraslow-cooling regime. As shown in Fig.~\ref{VortexSummary}, 
here the vortex density increases but without reaching expected value. 
Such behaviour was already discussed for other systems \cite{Ducci1999, Weir2013,Su2013} and attributed to vortex-antivortex annihilation. This explanation is consistent with our finding that the fall-off is largest at very slow cooling, when the time spent close to $T_\text{c}$ is large. It is also consistent with its absence in the flux-grown samples of Ref.~\cite{Chae2012}; our samples are grown using the floating zone method which we expect to have distinctly different stoichiometric modifications. We indeed observed that the mobility of vortices below $T_c$ can vary substantially among samples depending on the annealing gas atmosphere. 

We now analyze the pronounced trends with the chemistry across the $R$MnO$_3$ series that are clear in Figure \ref{VortexSummary}. In particular, the vortex density for a given cooling rate in the slow-cooling regime increases by three orders of magnitude from Tm to Dy, i.e.\ with decreasing $R^{3+}$ radius. In the Kibble-Zurek picture this is directly related to the microscopic parameters $\xi_0$ and $\tau_0$. Our first-principles calculations (see Table \ref{phyprop}) indicate that the mean-field value for $\xi_0$ is essentially the same in the four systems, as is the renormalized value $\tilde \xi_0$ ($\simeq 1.3 \xi_0$) in the fluctuation regime. Therefore, to obtain a factor $\sim 10^3$ in the vortex density in either the conventional mean-field or fluctuation-dominated Kibble-Zurek picture, $\tau_0$ would need to change by an unphysical factor of $\sim 10^6$. 
Strictly speaking, the renormalization of $\xi_0$ should be computed at the freeze-out temperature $T_*=[1 + (\tau_0/\tau_q)^{1/(1+z\nu)}]T_c$. However, this does not help us to reconcile the differences, since to increase the vortex density by the observed factor $10^3$, the freeze-out temperature would have to reach $T_* \sim  100 \, T_c \gtrsim 10^5$ K. This is again unphysical. We thus see that the vortex formation in the $R$MnO$_3$ systems displays quantitative features challenging the interpretation in terms of the original Kibble-Zurek mechanism. Specifically, the trend with chemistry does not fit with the standard scenario, even with the role of critical fluctuations beyond the mean-field description fully taken into account. 

Another trend in Fig.~\ref{VortexSummary} is the several-orders-of-magnitude shift of the cooling rate at which the departure from the 
Kibble-Zurek scaling occurs and which is not predicted by the calculated values for $Gi$ in Table \ref{phyprop}. 
It is striking, however, that the huge spread of the vortex density in the %Ginzburg-like 
Kibble-Zurek range is substantially reduced in the anti-Kibble-Zurek regime (from 1000 to 4), and in fact almost restores the approximate $R$-independence expected from Table \ref{phyprop}. 

At least in part, extrinsic factors may be responsible for these unexpected trends with chemistry. Temperature-dependent processes related to chemical or mechanical impurities could affect the vortex formation %below $T_c$ 
differently across the $R$MnO$_3$ series, simply because of the substantial increase of $T_c$ from Dy to Tm. In addition, the chemical and mechanical quality of the samples may change with $R$. DyMnO$_3$, in particular, is much harder to grow hexagonally than e.g. TmMnO$_3$ because of the closer proximity of the competing perovskite phase. In fact, we find that batches of the same material grown under slightly different conditions can show, at a specific cooling rate, a vortex-density spread up to factor four with  overall trends with cooling rate like in Fig.~\ref{fluct-to-mf} (see Supplementary Information). 

These factors can modify, in particular, the thermal conductivity of the samples. 
This conductivity is another ingredient limiting the overall thermalization of the system, and is such that the instantaneous temperature can additionally be position dependent across the sample. 
If this happens, the local transitions will not be simultaneous during the experiments. Instead, there will be a phase-transition front propagating at the speed $v_f=\frac{dx}{dt}|_{T(x,t)=T_c}$, which needs to be compared with the characteristic speed $v_c={\xi(\varepsilon)}/{\tau(\varepsilon)}=(\xi_0/\tau_0)|\varepsilon|^z$ of the order-parameter relaxation \cite{2013JPCM...25N4210D,Volovik:388144,1999PhRvL..82.4749D,Kibble:1996gt}. 
Thus, if the front propagates much slower than the order-parameter perturbations, the formation defects can be heavily suppressed in a composition-dependent fashion. 

Finally, we briefly discuss other possible origins of the anti-Kibble-Zurek behavior. The thermal conductivity could be one of them since, according to the above, 
%Moreover, 
a maximum in the vortex density can be expected at $v_f = v_c$, followed by quick decrease as the cooling rate (and hence $v_c$) increases.

On the other hand, the reversal from Kibble-Zurek to anti-Kibble-Zurek scaling has also been demonstrated in a recent Bose-Einstein-condensate transition experiment \cite{donadello16}. In that case, it has been ascribed to vortex-vortex repulsive interactions. Such interactions could also play a role in our $R$MnO$_3$ systems, which certainly display a rather dense vortex pattern in the crossover region (see Fig. \ref{VortexQuench}). 
In addition, there are additional time scales that can become relevant in the problem. 
The most obvious one is the time-scale needed to obtain a fully developed vortex from the initial seed. 
Indeed, by increasing the cooling rate, this process can be expected to be gradually suppressed --simply because the phonons will not have time to propagate the order-parameter perturbations \cite{zhang14}-- which will eventually lead to a decrease in the vortex density.

Thus the overall trend in Fig. \ref{VortexSummary} may still be largely determined by intrinsic properties, and hence directly related to the corresponding universality class. In this regard, the discrete nature of the symmetry breaking can play a crucial role, especially if, as has been repeatedly underlined (see e.g. \cite{rivers98,rivers00,rivers06}), the defect formation is decided after crossing the transition temperature. In fact, a generic feature of $Z_6$ models like Eq.\ \eqref{lfe} is the emergence of a second correlation length below $T_c$ associated with the phase of the order parameter supplementing that of its amplitude \cite{Lilienblum2015}. This, however, is missed in the usual Kibble-Zurek mechanism formulated for $U(1)$ models. This second correlation length diverges faster than the standard correlation length \cite{Lilienblum2015}. Thus, if the critical slowing down of the phase itself becomes a dominant process, the vortex density might become affected in a substantial way.

In summary, we have shown theoretically that the Kibble-Zurek picture inherently includes a transfer from the asymptotic scaling, where the system is well inside the Ginzburg region, to mean-field scaling, where fluctuations represent small perturbations. 
This transfer is general since all phase transitions, from the inter-atomic to the cosmological scale, are formally expected to undergo such a crossover. 
We matched our predictions with the experimental behavior in the series of hexagonal manganites, and showed that this crossover can be behind the striking reversal of the topological defect density as a function of the cooling rate (anti-Kibble-Zurek scaling). In addition, using density-functional-theory, we quantified the expected Kibble-Zurek behaviour across the $R$MnO$_3$ series. This quantification revealed the presence of sizeable chemical trends that call for a vital upgrade of the original Kibble-Zurek considerations for the systems in this class.

\small

\

\noindent{\bf Methods}

\noindent{\bf Sample preparation}
Samples were grown at ETH (DyMnO$_3$), PSI (YMnO$_3$%, DyMnO$_3$
) and Lawrence Berkeley National Laboratory (ErMnO$_3$) by the floating-zone technique as described in the supplementary information and elsewhere \cite{Roessli2005, Ivanov2006, Yan2015}. Rods were oriented by Laue diffraction and cut into $z$-oriented platelets of a few mm lateral size and a thickness of about 0.7~mm.

For minimizing sample-dependent drifts in measurements of the cooling-rate-dependent ferroelectric domain vortex density, all samples of a batch were pre-annealed under identical conditions by heating above $T_c$ for several hours (YMnO$_3$: 1400~K, DyMnO$_3$: 1270~K, ErMnO$_3$: 1470~K) and cooling through $T_c$ at a rate of 5~K/min. To reveal the intrinsic 3D bulk domain structure and suppress surface-dependent effects, samples were then thinned by at least 100~$\mu$m by lapping with Al$_2$O$_3$ powder. This was followed by etch-polishing with a silica slurry, revealing shiny surfaces with a rms-roughness below 1~nm before determining the vortex density by PFM. We found that this density showed systematic variations by a factor of about two, depending on the location of a sample in the original floating-zone rod. For our experiment we selected a sub-set of samples displaying the same vortex density after the pre-annealing within the statistical error of the vortex count. 

The quench experiments were performed like the pre-annealing experiments but with a dwell time above $T_c$ of few tens of minutes and varying cooling rate through $T_c$. In order to avoid accumulation of chemical drift occurring during the quench cycles, each data point was gained from a different specimen of the preselected set.

\noindent{\bf First-principles calculations}
For our density functional calculations we use the projector-augmented wave method as implemented in the abinit code \cite{Gonze:2002br,Gonze:2005em,Torrent:2008jw,Amadon:2008ia}. We use a plane-wave cutoff of 30 Ry and a $6\times 6\times 2$ k-point grid. To take into account correlation effects on the Mn atoms we use the LDA+U method within the fully localized limit method as introduced by Lichtenstein et al.~\cite{Perdew:1992ee, Liechtenstein:1995ip}. 
We choose a value of $U$ of 8~eV and $J=0.88$ eV. For all our calculations we adopt an A-type magnetic ordering of the Mn ions and freeze the rare-earth f electrons in the pseudopotential cores. To extract the parameters in the Landau free energy, Eqn.~\ref{lfe}, we first fully relax the $P6_3/mmc$ structure to an accuracy of $10^{-6}$ Ry/Bohr. We find unit cell volumes of 364.73 \AA$^3$, 364.42 \AA$^3$, 367.50 \AA$^3$ and 353.00 \AA$^3$ for YMnO$_3$, DyMnO$3$, ErMnO$_3$ and TmMnO$_3$ respectively. We then calculate the force constants using the finite displacement method and extract the eigenvectors of the force constant matrix. We then gradually freeze in the eigenvector of the unstable $K_3$ mode and fit a sixth order polynomial to extract the $a_0$ and $b$ terms. To calculate the gradient term $s$ we exploit the fact that in $q$-space $s(\nabla Q)^2$ reduces to $sq^2|Q_{\mathbf q}|^2$ and $s$ can then be obtained by fitting a parabola to the corresponding branch of the force constant dispersion as shown in \cite{Artyukhin2014}. We note that the values of the parameters are dependent on the choice of exchange-correlation functional because $a$ scales quadratically and $b$ to the fourth power with the lattice constant. The trends across the series, however, are robust to the computational details.

\

\noindent{\bf Acknowledgements}\\
\noindent We thank A. Varlamov for carefully checking Eq. \eqref{renormalized-xi0}, Thomas Weber for support concerning powder x-ray diffraction, which was performed at the X-Ray Service Platform of the Department of Materials of the ETH Zurich, 
and Thomas Lottermoser and Andrea Scaramucci for fruitful discussions. 
This work was supported by the ERC Advanced Grant Nr. 291151 CCICO. Computational resources were provided by ETHZ and by a grant from the Swiss National Supercomputing Centre (CSCS) under project ID p504.

\noindent{\bf Author contributions}\\
M.F. and N.S. initiated and coordinated the project. 
K.C., E.P. Z.Y., E.B., D.M., and F.L. provided the samples. M.L. performed the quenching experiments and determined the vortex density from the subsequent PFM measurements. Q.M. performed the DFT calculations and developed the KZ theoretical aspects together with A.C. Q.M., M.L., S.G., E.K.H.S., N.S., M.F., and A.C. discussed the results and their interpretation. Q.M., M.L., N.S., M.F., and A.C. wrote the manuscript.\\
$^{\ast}$These authors contributed equally to the manuscript.

\noindent{\bf Competing financial interests}\\
The authors declare no competing financial interests.

\end{document}